# The Fundamentals of Policy Crowdsourcing

By

John Prpić, Araz Taeihagh[*], and James Melton

[*]Corresponding Author: araz.taeihagh@new.oxon.org

**The Fundamentals of Policy Crowdsourcing**

John Prpić, Faculty of Business Administration, Technology and Social Sciences – Lulea University of Technology

Araz Taeihagh, School of Social Sciences, Singapore Management University, Singapore [araz.taeihagh@new.oxon.org]

James Melton, Department of Business Information Systems, Central Michigan University, Mount Pleasant, Michigan, US

What is the state of the research on crowdsourcing for policy making? This article begins to answer this question by collecting, categorizing, and situating an extensive body of the extant research investigating policy crowdsourcing, within a new framework built on fundamental typologies from each field. We first define seven universal characteristics of the three general crowdsourcing techniques (virtual labor markets, tournament crowdsourcing, open collaboration), to examine the relative trade-offs of each modality. We then compare these three types of crowdsourcing to the different stages of the policy cycle, in order to situate the literature spanning both domains. We finally discuss research trends in crowdsourcing for public policy, and highlight the research gaps and overlaps in the literature.

# Introduction

Crowdsourcing (Howe 2006, 2008; Brabham 2008) involves organizations using IT to engage crowds comprised of groups and individuals for the purpose of completing tasks, solving problems, or generating ideas. In the last decade, many organizations have turned to crowdsourcing to engage with consumers, accelerate their innovation cycles, to search for new ideas, and to create knowledge (Afuah & Tucci, 2012; Majchrzak et al. 2012; Bayus 2013; Brabham 2013a). As crowdsourcing has become an increasingly popular method for business organizations to gather IT-mediated input from individuals, the phenomenon has also spread to non-commercial contexts too. Recently, crowdsourcing has begun to be applied to different aspects of policy making, for example in the transportation (Nash 2009) and urban planning domains (Seltzer & Mahmoudi 2013). Yet, despite the advancing use of crowdsourcing in general, and its recent application in policy contexts, to our knowledge, research has yet to emerge that systematically investigates both domains simultaneously. This article is an attempt to address this salient gap in our knowledge. To do so, we introduce two fundamental typologies, one each from the crowdsourcing and policy literatures, which we then merge to form a new systematic framework suitable to address all applications of policy crowdsourcing. We then employ our new framework to situate and organize an extensive body of the extant literature on policy crowdsourcing. We thereby illustrate the current trends, highlight the research gaps, outline the trade-offs, and provide a systematic approach to investigating policy crowdsourcing.

In the next section we introduce the three different types of crowdsourcing, before introducing the policy cycle. We then detail the trade-offs inherent to the different types of crowdsourcing, illustrating the potentials and constraints along seven independent factors relevant to crowdsourcing techniques. Thereafter, we merge the frameworks to form a new framework, allowing us to organize and situate an extensive body of literature on policy crowdsourcing, therein highlighting the apparent gaps and overlaps in the extant research. Finally, we discuss the implications of our study for the research and policy practitioner communities, before concluding with a summary of the article's contributions.

## Crowdsourcing

Crowdsourcing is an IT-mediated problem-solving, idea-generation, and production model that leverages the dispersed knowledge of groups and individuals to produce heterogeneous resources for organizations (Hayek 1945; Brabham 2008; Prpić & Shukla 2013). Problem solving, idea generation and production are sourced from crowds through the means of IT, such as via virtual labor markets (Horton 2010; Horton & Chilton 2010; Wolfson & Lease 2011; Irani & Silberman 2013), open collaboration (Crump 2011; Small 2012; Adi, Erickson & Lilleker 2014) or through tournament-based competitions (Piller & Walcher 2006; Blohm et al. 2011; Schweitzer et al. 2012). As an overall approach to engaging dispersed knowledge through IT, crowdsourcing processes serve to blend the efficiency and control of traditional, top-down managed processes, with the benefits of bottom-up innovation and creativity (Brabham 2008; Howe 2006, 2008). Organizations can launch crowdsourcing initiatives on their own in-house platforms, therein seeking to coalesce a proprietary crowd, as commercial organizations such as Dell (IdeaStorm), Quirky, and Starbucks (MyStarbucks) illustrate, or an organization can commission crowdsourcing intermediaries to provide the requisite IT means and a "built-in" crowd, as a paid service (Bayus 2013) on platforms like eYeka, Kaggle, and Innocentive.

Though crowdsourcing phenomena continue to evolve in form and function, the crowdsourcing literature has recently begun to coalesce around three distinct IT-mediated forms: virtual labor markets, crowdsourcing tournaments, and open collaboration (Estellés-Arolas & González-Ladrón-de-Guevara 2012; de Vreede et al. 2013; Prpić, Jackson & Nguyen 2014). In the following subsections, we will focus on each distinct type in turn.

*Virtual Labor Marketplaces (VLMs)*

A virtual labor marketplace (VLM) is an IT-mediated market for spot labor, typified by endeavors like Amazon's M-Turk and Crowdflower, where individuals and organizations can agree to execute work in exchange for monetary compensation (Horton 2010; Horton & Chilton 2010; Wolfson & Lease 2011; Irani & Silberman 2013). These endeavors are generally thought to exemplify the 'production model' aspect of crowdsourcing (Brabham 2008), where workers undertake microtasks for pay. Microtasks, such as the translation of documents, the tagging of photos, and transcribing audio (Narula et al. 2011), are generally considered to represent forms of human computation (Iperiotis & Paritosh 2011; Michelucci 2013), where human intelligence is asked to undertake tasks that are not currently achievable through artificial intelligence.

The size of the overall crowd available at these VLMs is massive, with Crowdflower for example, having over five million potential laborers available. Microtasking through VLMs can therefore be completed rapidly (if need be) through the massively parallel scale available on such platforms. The participants in these VLM crowds generally undertake tasks independent of one another, and thus do not form official groups, or work as teams, through the intermediary platforms. Further, the laborers in VLMs are largely anonymous (Lease et al. 2012) with respect to their offline identities.

*Tournament Crowdsourcing (TC)*

A separate form of crowdsourcing is known as tournament crowdsourcing (TC) or ideas competitions (Piller & Walcher 2006; Blohm et al. 2011; Schweitzer et al. 2012). In TC, organizations post their problems to IT-mediated crowds on platforms such as Innocentive, Eyeka, and Kaggle (Afuah & Tucci 2012) or through in-house platforms such as Challenge.gov (Brabham 2013b). These platforms generally attract and maintain more or less specialized crowds premised on the platform's specific focus; for example, Eyeka's crowd creates advertising collateral for brands, while the crowd at Kaggle focuses on data science solutions (Ben Taieb & Hyndman 2013; Roth & Kimani 2013). When applied to innovation, these platforms have been termed open innovation platforms (Sawhney et al. 2003), and represent both the idea generation and problem solving aspects of crowdsourcing (Brabham 2008; Morgan & Wang 2010).

The numbers of participants at these sites is smaller than at VLMs (for example, Kaggle has approximately 140,000 available, compared with the millions on Crowdflower), and the individual participants can choose not to be anonymous at these sites. Fixed amounts of prize money, and fixed numbers of prizes, are generally offered to the crowd for the best solutions submitted, and prizes can range from a few hundred dollars to a million dollars or more.[†] Some TC intermediaries require that their crowds submit independent solutions to competitions (e.g., eYeka), while others such as TopCoder allow or even encourage team formation and, thus, within-crowd collaboration in competitions.

*Open Collaboration (OCs)*

---

[†] http://www.innocentive.com/files/node/casestudy/case-study-prize4life.pdf.

In the open collaboration model of crowdsourcing, organizations post their problems or opportunities to the public at large through IT (Crump 2011; Small 2012; Adi, Erickson & Lilleker 2014). Contributions from the crowds in these endeavors are voluntary and thus do not generally entail monetary exchange. Starting an enterprise wiki (Jackson & Klobas 2013) or using social media (Kietzmann et al. 2011) like Facebook and Twitter (Gruzd & Roy 2014; Sutton, et al. 2014) to garner contributions, are prime examples of this type of crowdsourcing.

The scale of the crowds available to these types of endeavors can vary significantly depending on the reach and engagement of the IT used, and the efficacy of the 'open call' for volunteers. For example, as of March 2015, Twitter had approximately 288 million registered users and though this crowd is immense, there is little to guarantee the attention of any significant subset of the contributors when using Twitter to crowdsource. It is also important to note in respect to OCs that the crowds in these endeavors are much less constrained with respect to self-organization (Prpić & Shukla 2013) than the other two types of crowdsourcing. What this means for organizations is that the individuals in these OC crowds, by virtue of their ready access to the same tools that the organizations are using, have the opportunity to alter or amplify the agenda of organizational OCs through their own personal IT-mediated networks.

## The Policy Cycle

Jenkins (1978) defined public policy as "a set of interrelated decisions taken by a political actor or group of actors concerning the selection of goals and the means of achieving them within a specified situation where those decisions should, in principle, be within the power of those actors to achieve." As such, a policy can be construed as a set of effective and acceptable courses of action implemented to reach explicit goals (Bridgman & Davis 2004). Implicit within this view is the assumption that policy makers are rational, though this assumption has been vigorously debated at times (Kingdon 1984; Stone 2002).

An early proponent of simplifying policymaking by breaking it down to interrelated stages for the purpose of analysis was Lasswell (1956). This systemic analysis for understanding and explaining political systems served to convert inputs such as political demands and political support to outputs in the form of decisions and actions (Easton 1979), an idea that was later extended to policies by Palmer (1997).

Various attempts at classification of the different stages of the policy cycle have been carried out over the years. In this article, based on the efforts of Stone (1988) and Howlett et al. (1995), the policy cycle is seen as a sequence of steps in which agenda setting, problem definition, policy design, policy implementation, policy enforcement, and policy evaluations are carried out in an iterative manner (Figure 1).

Figure 1. The Policy Cycle.

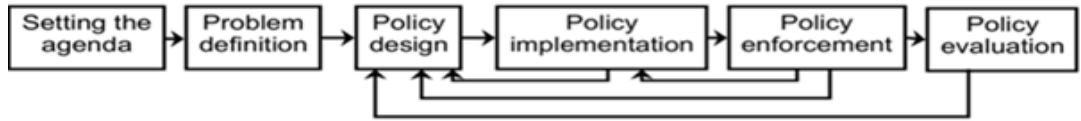

## Comparison of Crowdsourcing Techniques

Prpić, Taeihagh & Melton (2014a) compared the three types of crowdsourcing discussed above across three universal dimensions—cost to implement, anonymity of individuals in the crowd, and scale of the crowd—using three-point estimates for each characteristic where possible. Here, we build upon that work by extending the comparison to include four additional common characteristics to illustrate the stable and relative differences between the various forms of crowdsourcing more comprehensively. This set of characteristics reflects a minimum and general consensus extracted from the literature (de Vreede et al. 2013; Estellés-Arolas & Ladrón-de-Guevara 2012; Prpić & Shukla, 2013, 2014; Prpić, Shukla, Kietzmann, & McCarthy 2015) and does not represent either an exhaustive set of characteristics nor a perfectly independent categorization, since many crowdsourcing applications are hybrids of a sort, known to mix elements and features of the 'pure play' forms that have thus far emerged.

The dimensions used for comparison of all crowdsourcing techniques will be discussed in turn. These are: cost; anonymity; scale of the crowd; IT structure; time required to implement; task magnitude; and reliability of the crowd.

*Cost*

The cost dimension refers to the typical cash outflows for an organization when implementing a crowdsourcing technique for any purpose. Open Collaboration crowdsourcing techniques, such as the use of Twitter, Reddit, Facebook, or a Wiki (etc.), are essentially free for the implementer in terms of direct cash outlays for the use of the IT artifact, while TC and VLMs necessitate explicit cash outlays for their use. TC endeavors are generally fixed-cost cash outlays, where the organization sets the number of prizes in a tournament and the value per prize, ahead of launching the competition. VLMs on the other hand, necessitate variable-cost cash outlays on a per completed task basis. Since TC and VLMs are generally accessed through third party intermediation services, variable-cost cash outflows are also needed to compensate the service provider.

*Anonymity*

The anonymity dimension of our comparison refers to whether the participants in the crowds at each of the three generalized crowdsourcing types are anonymous (or not) with respect to their offline identity. In some cases, predominantly in forms of Open Collaboration (such as in Google+, Facebook, Twitter, or enterprise Wikis), the online and offline identity are generally twinned. In VLMs, however, there is essentially a form of 'methodological anonymity' found in all the intermediary platforms providing these services, where crowd-workers are identified only by unique numeric identifiers (see Lease, et al. 2013, for an important exception). At the TC intermediaries, anonymity is 'medium' in our estimation, since generally these platforms do not necessitate a matching of online and offline identity, though some intermediaries give crowd members the choice to use a pseudonym or their offline identity. At some TC sites, such as Kaggle or Innocentive, there may actually be strong incentives for high-performers to give up their anonymity, so that their excellent performances can bolster their offline career prospects. Taken altogether, the relative anonymity of crowd participants is important since anonymity is one method of maintaining privacy. Those organizations that are concerned with maintaining privacy for legal, ethical, or moral reasons (e.g., researchers and health care organizations) or are mandated to do so (e.g., government institutions) will need to consider the liability that the different forms of crowdsourcing anonymity entail (Wolfson & Lease 2011).

*Scale of Crowd Size*

The size of crowds available to organizations implementing one of the generalized forms of crowdsourcing varies for each form. VLMs like Crowdflower boast over five million members available for an organization to access, while the most successful TC intermediaries like Kaggle, eYeka, and Innocentive boast crowds of hundreds of thousands of members. In effect, there may be something approaching or surpassing an order of magnitude of difference between VLM and TC crowds in general. With respect to Open Collaboration, the situation varies significantly with applications like Twitter and Facebook having hundreds of millions of members, an order of magnitude or more than the largest VLMs. Other forms of open collaboration, such as enterprise wikis, will of course vary in crowd size only to the extent of the size of the firm itself, being implemented similarly whether with handfuls of individuals forming a crowd in SMEs or Not-For-Profits, to tens of thousands of individuals in the largest firms. The crowd size differences outlined here are important for organizations to understand when choosing the form of crowdsourcing to implement, since the scale of the crowd represents a maximum limit to the number of potential contributors in a crowd, and thus is a potential constraint to the quantity of crowd contributions, and the speed by which the desired contributions can be gathered.

*IT Structure*

The IT structure of each crowdsourcing type can be found to exist in either episodic or collaborative form, premised on the interface of the IT used to engage a crowd (Prpić & Shukla 2013, 2014; Prpić, Shukla, Kietzmann, & McCarthy, 2015). Episodic forms of crowd-IT do not require that crowd members interact with one another through the IT in order for resources to be derived from the crowd (see Google's reCAPTCHA for example). The reverse is true of collaborative forms of crowd-IT, where participants have to interact with one another through the IT, for the organization to derive resources from the crowd (such as enterprise wikis for example). This distinction is crucial, given that all forms of crowdsourcing are IT-mediated phenomena. Said another way, in episodic forms of crowd-IT, social capital does not need to exist, be created, or maintained through the IT for crowd-derived resources to be created. The reverse is true for collaborative forms of crowd-IT.

In our comparison of crowdsourcing types, we find that VLMs are found to use episodic IT structures and OCs are found to generally use collaborative IT structures, while TC varies in this respect, where examples or elements of both forms of IT structure can be found to exist. Organizations considering implementing a type of crowdsourcing, or endeavoring to build their own, must give serious consideration to these matters, given that the IT structure determines both the interaction between the organization and the crowd, and the potential interaction of the crowd-participants with each other.

*Time to Implement the Form of Crowdsourcing*

The time required to implement a particular crowdsourcing technique varies considerably amongst the available options. In our estimation, VLMs necessitate the least amount of lead time to begin gathering crowd contributions, given the vast amount of on-demand labor available at all times at these platforms, and that one can join a VLM and begin receiving contributions from the crowd within minutes. On the other hand, using TC intermediaries like eYeka, Innocentive, or Kaggle is a much more involved process, where the sponsoring organization generally works with the competition-hosting intermediary to design the contest, and the prize and money distribution, before any crowd members become involved. Further, competitions necessitate choosing an appropriate duration for the contest itself, generally ranging from a couple of weeks to many months or more. Similarly, at the end of the contest, the winning submissions must be selected from a sometimes vast range of competition entries received. Open Collaboration, on the other hand, varies in this respect, and is premised on whether an organization already has the collaborative system in place or not. For example, if an organization has invested in developing a highly followed Twitter or Facebook network over time, then the time to receive crowd contributions can be almost nil. On the other hand, if such a presence needs to be built from scratch, the time to contribution from the crowd can be very long indeed. Further, given that OC contributions are generally voluntary in nature, there is no guarantee that contributions ever manifest at all.

*Task Magnitude*

Task magnitude refers to the size and complexity of the tasks asked of and received from the crowd. VLMs are generally considered to elicit forms of human computation in the form of microtasks; TC endeavors are generally thought to elicit complete solutions to specific problems posed; while OCs can traverse the entire spectrum from microtasks to complete solutions, depending on a particular implementation. Task magnitude is an important consideration for organizations both within a form of crowdsourcing, and amongst the three forms taken together (Basak, Loni & Bozzon 2014; Nakatsu, Grossman & Iacovou 2014; Wagner & Suh 2013). In VLMs, task size considerations are important, given the massive parallel scale with which tasks can be undertaken, and because it is necessary to place an attractive price for each task into the market to attract crowd members to execute them. On the other hand, in TC, task magnitude is essentially offered at the solution level. In other words, organizations ask for and receive fully formed solutions to the entire problem that they offer up for the competition. In respect to OC task magnitude, broad variation exists, spanning the spectrum from microtask to complete solutions. If an organization, for example, uses Twitter or Facebook to gather ideas from a crowd, such crowd inputs would be closer to a microtask in a VLM, especially given the limitations of these platforms (i.e., 140 characters in Twitter). On the other hand, the use of an enterprise wiki in an organization is expected to accrue and evolve over time through crowd contributions, and may approach something resembling a permanent, yet adjustable, knowledge repository with relatively complete solutions. In all cases of crowdsourcing, an organization will need to undertake some pre-task preparation and some form of post-task processing of crowd contributions, in order to generate the desired value, and task magnitude is central to these concerns.

*Reliability of the Crowd*

The reliability of the crowd refers to the general consistency of each form of crowdsourcing to supply the desired inputs for an organization. TC is considered to provide the highest level of reliability, given that said solutions are something approaching complete, and that such crowds are accessible on demand. VLMs are considered to be of 'medium' reliability, given that these crowds are available on demand just like TC, though somewhat less reliable given the uncertainty around the price/time equation in these markets, and the known presence of malicious or fraudulent workers (Wang et al. 2014). In OCs, once more, we see variability along this dimension relative to the specifics of the form of OC chosen. For example, if an organization uses Reddit, there is little to guarantee that any significant subset of the approximately three million Redditors will engage with the organization's effort. On the other hand, when it comes to an enterprise wiki for example, an organization may be able to enforce or incentivize its entire employee base to participate in short order, therefore considerably increasing the reliability of such an OC crowd.

*Summary*

In this section we have aimed to provide a relative comparison of the three modes of crowdsourcing in a generalized form by unpacking the modes amongst the universal characteristics identified (see Table 1). In doing so, we highlight some generalized trade-offs faced by organizations in their implementation of crowdsourcing techniques for any purpose. Our categorizations and assigned values are neither exhaustive nor definitive; rather, we have aimed to usefully extend the crowdsourcing literature in this respect.

**Table 1. Comparison of Common Characteristics of Crowdsourcing Techniques.**

| **Common Characteristics** ------------------- **Modes of Crowdsourcing** | Cost | Anonymity | Scale of Crowd | IT Structure | Time required to implement | Task magnitude | Reliability of the Crowd |
|---|---|---|---|---|---|---|---|
| Virtual Labor-Markets | **Variable** | **High** | **High** | **Episodic** | **Low** | **Simple** | **Medium** |
| Tournament-Based Collaboration | **Fixed** | **Medium** | **Medium** | **Variable** | **Medium** | **Complex** | **High** |
| Open Collaboration | **Free** | **Variable** | **Variable** | **Collaborative** | **Variable** | **Variable** | **Variable** |

## Policy Crowdsourcing Framework

In this section, we combine the preceding analyses to create an overarching policy crowdsourcing framework that includes the different types of crowdsourcing techniques merged with the various stages of the policy cycle. We then populate the resulting table (Table 2) with an extensive body of literature in the policy crowdsourcing domain.

For our analysis, we bound the policy crowdsourcing domain by limiting it to research that investigates crowdsourcing phenomena implemented or discussed in the purview of public administration, at any stage of the policy cycle. Public Administration can occur at any level of government, though in our view it explicitly excludes the use of crowdsourcing techniques for 'vote-getting' by politicians (Jungherr 2014). Thus, while research on a politician using Facebook or Twitter to gather supporters would be excluded from our dataset (Hemphill, Otterbacher, & Shapiro 2013), research investigating the practice of a member of a legislature using a wiki page or Twitter to solicit ideas relevant to legislation would be included (Mainka, Hartmann, Stock, & Peters 2014).

Similarly, it's important to note that the use of some forms of social media strictly for 'informing' or 'service delivery' purposes, such as in many e-government initiatives (Small 2012; Criado, Sandoval-Almazan, & Gil-Garcia 2013), are also excluded from our literature dataset. Although all forms of social media can be used for Open Collaboration crowdsourcing, the use of social media does not in itself guarantee that crowdsourcing is occurring. For Open Collaboration to occur, like all forms of crowdsourcing, explicit resources of some sort (data, information, knowledge, money, work, etc.) need to be generated from the crowd through the IT used.

To increase the focus and utility of our framing exercise, we exclude the large and growing literature on crisis mapping (see for example Norheim-Hagtun & Meier 2010; Birregah et al. 2012; Meier 2012; Ziemke 2012; Bott et al. 2014), and the burgeoning literature on crowdfunding (see for example Aitamurto 2011; Wheat et al. 2013) where they may pertain to public administration, since these literatures are already relatively freestanding.

For the sake of further clarity and thoroughness, we now delimit crowdsourcing from other concepts in use in the management and governance literatures that may share some elements in common. In the innovation realm, research on innovation networks (von Hippel 2005a; 2005b), co-creation (Prahalad &Ramaswamy 2004; Zwass 2010; Voorberg, Bekkers & Tummers 2014) and open innovation (Chesbrough 2003; Asakawa, Song & Kim 2014) has emerged within the last decade. Similarly, in what might be considered more of a governance context (Howlett & Lindquist 2007), commons-based peer production (Benkler & Nissenbaum 2006; Hill & Monroy-Hernández 2012), mass collaboration (Panchal & Fathianathan 2008; Tapscott & Williams 2008; Doan, Ramakrishnan & Halevy 2010; Tkacz 2010), social computing (Ala-Mutka et al. 2009; Punie, Misuraca & Osimo 2009), civic, public, and citizen participation (Newman et al. 2004; Anduiza, Cantijoch & Gallego 2009; AbouAssi, Nabatchi & Antoun 2013; Coelho 2014; Le Dantec 2014), deliberative democracy (Himmelroos & Christensen 2014; Nabatchi 2014), and e-government (Criado, Sandoval-Almazan, & Gil-Garcia, 2013; Bertot, Jaeger & Grimes 2010, 2012) each have burgeoning literatures associated with the subjects.

Although it is beyond the scope of this article to compare the detailed similarities and differences of each of these concepts with crowdsourcing, as others have begun to do (Quinn & Bederson 2011), each of these concepts taken severally illustrate substantial differences from crowdsourcing on one or more of the following dimensions:

- Mixing offline and online phenomena (i.e. citizen participation, open innovation, innovation networks)
- Different focus or level of analysis (i.e. co-creation, peer production, deliberative democracy, e-government)
- Abstraction from the IT artifact (i.e. IT artifacts in crowdsourcing are explicitly tied to the specific phenomena itself)

In short, the seven universal characteristics of crowdsourcing that we draw upon here, as a corpus, do not pertain to any of these other concepts systematically.

*Data Collection*

Data collection through the use of secondary archival sources such as search engines,[‡] alerts, social media, webpages, the general press, blogs, etc. began in December 2013.

---

[‡] To give the reader some indication of the depth of the search for literature, in early 2015 a Google Scholar search was undertaken for "crowdsourcing"+"policy", and 27 pages of the search

Over time, 189 pieces of literature were collected, of which 83 remain relevant for this work after data filtering. The 83 articles captured in this analysis are limited to ones that research the use of crowdsourcing in one or more areas of the policy cycle, and all the research works in question self-identify in this manner, to greater or lesser extent. The literature included journal articles, peer-reviewed conference articles, book chapters, theses, technical reports, and books. Implementing the two fundamental typologies that we use to form our policy crowdsourcing framework, all 83 pieces of literature were coded by the research team to fall into one of the three forms of crowdsourcing and one stage of the policy cycle.

**Table 2. Policy Crowdsourcing Framework with Relevant Literature.**

|  | **VLMs** (E.G. M-Turk, Crowdflower) | **Tournaments** (E.G. Innocentive, Challenge.gov) | **Open Collaboration** (E.G. Twitter, Wikipedia, Ushahidi) |
|---|---|---|---|
| **Agenda Setting** |  | Brabham (2012b) Brabham (2013b) Desouza & Krishnamurthy (2014) | Osimo (2008) Ala-Mutka et al. (2009) Punie et al. (2009) Jungherr & Jürgens (2010) Nam (2010) Lackaff & Grímsson, (2011) Lindner & Riehm (2011) Linders & Wilson (2011) Aitamurto (2012) Bonson et al. (2012) Charalabidis et al. (2012) Bani (2012) Bua (2012) Kriplean et al. (2012) Linders (2012) Liu (2012) Christensen et al. (2014) Clark et al. (2013) Cupido & Ophoff (2014) Crawford et al. (2014) Garcia et al. (2015) Heikka (2014) Mainka et al. (2014) Shahsavarani (2014) Spiliotopoulou et al. (2014) Von Lucke (2014) |

results (ie the first 540 results, of the overall 19,200 results) returned (including patents and citations) were reviewed entry by entry by the research team until search saturation was achieved. In this context, our heuristic for search saturation was achieved when the review of the search results revealed only a duplication of known literature and a lack of new relevant results for three consecutive search pages.

| | | | |
|---|---|---|---|
| | | | Gellers (2015)<br>Johnston (2015) |
| **Problem Definition** | | Basto et al. (2010)<br>Mergel & Desouza (2013) | Chun et al. (2010)<br>Mergel (2012)<br>Nam (2012)<br>Ferro et al. (2013)<br>Buntaine et al. (2014)<br>Loukis et al. (2014)<br>Khan et al. (2014)<br>Offenhuber (2014)<br>Ramos (2014) |
| **Policy Design** | Prpić et al. (2014b) | Federal Prize Authority (2014)<br>Mergel et al. (2014) | Brito (2008)<br>Nash (2009)<br>Basto et al. (2010)<br>Bicquelet & Weale (2011)<br>Fung & Warren (2011)<br>Koch et al. (2011)<br>Warner (2011)<br>Chun & Cho (2012)<br>Lee & Kwak (2012)<br>Seltzer & Mahmoudi (2012)<br>Stottlemyre & Stottlemyre (2012)<br>Haklay et al. (2014)<br>Matei & Irimia (2014)<br>Moss & Coleman (2014)<br>Nelimarkka et al. (2014)<br>Osella )2014)<br>Raffl (2014)<br>Tambouris et al. (2014)<br>Taudes & Leo (2014)<br>Aitamurto & Landemore (2015)<br>Bertone et al. (2015)<br>May et al. (2015) |
| **Policy Implementation** | | Brabham (2012a) | Brabham (2013c)<br>Leeman et al. (2014)<br>Panagiotopoulos et al. (2014) |
| **Policy Enforcement** | Kim et al. (2013) | | Noveck (2009)<br>Ghafele (2011)<br>Bailard & Livingston (2014)<br>Hellstrom (2015) |
| **Policy Evaluation** | | | Benkler et al. (2013)<br>Franklin et al. (2013)<br>Balagapo et al. (2014)<br>Lee et al. (2014)<br>Schintler & Kulkarni (2014)<br>Kim et al. (2015)<br>Lodge & Wegrich (2015) |

*Framework Analysis*

From Table 2 it is immediately evident that the academic research on policy crowdsourcing is relatively sparse (though seemingly growing rather rapidly), and a large portion of the potential space that might be covered lacks research. Of the sparse research that does exist, the vast majority is focused upon OC crowdsourcing applications for policy purposes, while VLMs and TC have been relatively ignored. Perhaps not surprisingly, almost all of the research in respect to policy crowdsourcing has emerged within the last five years, with much of the literature being more recent than that. This seems to signal a growing application of crowdsourcing for policy, given that most of the research is premised upon investigations and discussions of policy practitioner implementations.

As it stands at this point in time, OC holds the predominant share of research on policy crowdsourcing, and this fact bears further fine-grained investigation. Is OC simply the crowdsourcing application most suited for policy concerns? Or are there other factors at play? In contrast, VLMs are barely represented in the research, and the reasons for this are not clear. It seems that the possible use of VLMs for policy has not caught on among policymakers, leading to lower levels of academic research on the topic. Or it may be that VLMs are being used, but that this use is not well publicized, as opposed to TC and OC, which benefit from being publicized, and are, by nature, more observable to outsiders. Further, it may be that ethics or national data sovereignty issues (Irion 2012) are more of a concern and therefore more of a hurdle to implementing VLMs for policy. In any case, at the moment, VLMs are a largely untapped resource in this area. Similarly, TC is scarcely represented in the literature, which is somewhat surprising, given the relatively high-profile success of Challenge.gov (Brabham, 2013b) and of open innovation platforms in non-policy domains in general.

In terms of the research methods employed in the literature we reviewed, case studies (Yin 2014) and other phenomenon-based methods (Miles & Huberman 1994, Flick 2002) are the most common approaches to investigation and data collection. These methods have the benefit of being grounded in real-world conditions; however, they typically require considerable amounts of time to undertake, can be resource-intensive, and are generally limited in terms of generalizability. On the other hand, experiments (Trochim 2005) that utilize the various crowds and applications available have been scarcely used to date. With the relatively low costs involved and the low barriers to entry with crowdsourcing, such experiments should be considered as a viable means to address the research gaps we have identified in the application of crowdsourcing to the policy cycle.

# Conclusion

We began this work with the question: what is the state of crowdsourcing in policy making? Our review of the literature on crowdsourcing for policy identified 83 works recently published on the subject, which indicates that crowdsourcing is already being used in different stages of the policy cycle and that numerous researchers have deemed these efforts important and/or interesting enough to thoroughly study these instantiations. It may be that this literature may serve as a bellwether of sorts, indicating a new and rapidly emerging field of interdisciplinary social science inquiry. In addition to collecting an extensive body of the current research in one review, we situate and organize the existing research in our systematic policy crowdsourcing framework. In doing so, we learn that research investigating all three forms of crowdsourcing for policy already exists, though it is very far from evenly distributed across all the stages of the policy cycle or the forms of crowdsourcing.

Similarly, our situation of the extant literature in the broader crowdsourcing and policy cycle frameworks readily indicates numerous gaps where no research exists at all. The reasons for this are unclear, though these voids represent useful research opportunities for future work. On the other hand, it may be that these apparent voids represent a lack of real-world application, indicating that TC and VLMs are not being used for crowdsourcing those stages of the policy cycle. If this is at least partially true, then these gaps also represent opportunities for policy-makers and researchers alike to pioneer such efforts. Given the demonstrated value of all the crowdsourcing techniques in other domains, it may be that these untapped potentials, severally or in combination, may represent powerful avenues to crowdsource different stages of the policy cycle.

In addition to the above contributions, we have systematically brought the crowdsourcing literature into the realm of policy, therein helping to define the outlines of a new, emerging, and socially salient context for crowdsourcing application. It may be that such an approach will encourage the relatively large number of researchers studying crowdsourcing to take this new context more seriously, perhaps helping to import more crowdsourcing expertise into the policy domain. Further, we explicitly extend the general crowdsourcing literature (irrespective of the policy context) by highlighting and detailing seven universal dimensions by which the three generalized forms of crowdsourcing can be usefully compared. We hope that the findings of this work are useful and accessible to both the research and policy practitioner communities, and that they increase the understanding and appreciation of the immense potential of policy crowdsourcing for experts in both the crowdsourcing and policy domains.

Meier, P. (2012). Crisis mapping in action: How open source software and global volunteer networks are changing the world, one map at a time. Journal of Map & Geography Libraries, 8(2), 89-100.

Mergel, I. (2012). Distributed democracy: Seeclickfix.com for crowdsourced issue reporting. Social Science Research Network.

Mergel, I., & Desouza, K. C. (2013). Implementing open innovation in the public sector: The case of challenge. gov. Public administration review, 73(6), 882-890.

Mergel, I., Bretschneider, S. I., Louis, C., & Smith, J. (2014). The Challenges of Challenge. Gov: Adopting Private Sector Business Innovations in the Federal Government. In System Sciences (HICSS), 2014 47th Hawaii International Conference on (pp. 2073-2082). IEEE.

Michelucci, P. (2013). Handbook of Human Computation. Springer New York.

Miles, M. & Huberman, A.M. (1994). Qualitative Data Analysis. Sage: Thousand Oaks, CA.

Morgan, J. & Wang, R. (2010). Tournaments for Ideas. California Management Review. 52(2), p.77–97.

Moss, G. & Coleman, S. (2014). Deliberative Manoeuvres in the Digital Darkness: e-Democracy Policy in the UK. The British Journal of Politics & International Relations, 16(3), 410-427.

Nabatchi, T. (2014). Deliberative Civic Engagement in Public Administration and Policy. Journal of Public Deliberation, 10(1), 21.

Nakatsu, R. T., Grossman, E. B., & Iacovou, C. L. (2014). A taxonomy of crowdsourcing based on task complexity. Journal of Information Science, 0165551514550140.

NAM, T. (2010). The wisdom of crowds in government 2.0: Information paradigm evolution toward wiki-government. AMCIS 2010 Proceedings. Paper 337.

Nam, T. (2012). Suggesting frameworks of citizen-sourcing via Government 2.0. Government Information Quarterly, 29(1), 12-20.

Narula, P., Gutheim P., Rolnitzky, D., Kulkarni, A. & Hartmann, B. (2011). MobileWorks: A Mobile Crowdsourcing Platform for Workers at the Bottom of the Pyramid. In Proceedings of HCOMP.

Nash, A. (2009). Web 2.0 Applications for Improving Public Participation in Transport Planning. In Transportation Research Board 89th Annual Meeting.